# SW# - GPU enabled exact alignments on genome scale

Korpar, Matija[1], Šikić, Mile[1,2*]

[1] University of Zagreb, Faculty of Electrical Engineering and Computing, Unska 3, HR 10000 Zagreb, Croatia

[2] Bioinformatics Institute, A*STAR, 30 Biopolis Street, #07-01 Matrix, 138671 Singapore

## ABSTRACT

**Summary:** Sequence alignment is one of the oldest and the most famous problems in bioinformatics. Even after 45 years, for one reason or another, this problem is still actual; current solutions are trade-offs between execution time, memory consumption and accuracy. We purpose SW#, a new CUDA GPU enabled and memory efficient implementation of dynamic programming algorithms for local alignment. In this implementation indels are treated using the affine gap model. Although there are other GPU implementations of the Smith-Waterman algorithm, SW# is the only publicly available implementation that can produce sequence alignments on genome-wide scale. For long sequences, our implementation is at least a few hundred times faster than a CPU version of the same algorithm.

**Availability:** http://complex.zesoi.fer.hr/SW.html

## 1 INTRODUCTION

Sequence alignments are fundamental in bioinformatics as they can provide information about unknown genes and proteins. They also have an important role in comparative sequence assembly (Flicek and Birney, 2009). In sequence alignments, both time and accuracy determines the successful performance of the method. The calculation time has acquired a high importance, mainly due to the growing amount of data coming out from the high throughput sequence initiative. On the other hand, obtaining the best quality results is at least as important.

Two classic algorithms for local alignment are Smith-Waterman (Smith and Waterman, 1981) and BLAST (Altschul *et al.*, 1990). The Smith-Waterman algorithm, an exact similarity searching method, aims to find the best local alignment between two sequences. BLAST (Basic Local Alignment Search Tool), an heuristic approach, identifies local alignments using two steps. First, it finds short matches and then creates local alignments around these initial matches. BLAST, being a heuristic algorithm, does not guarantee an optimal alignment of two sequences. It was shown on protein sequence databases that the Smith-Waterman algorithm performs better than BLAST (Pearson, 1995; Shpaer *et al.*, 1996). However, BLAST runs faster and requires less space (memory) consumption. This advantage becomes of high relevance for the alignment of long DNA sequences.

Although the original Smith-Waterman algorithm has quadratic time and space complexity, it is possible to reduce the space complexity to linear by increasing required computational time about two to three folds (Hirschberg, 1975; Myers and Miller, 1988; Chao *et al.*, 1994). In spite of the memory consumption improvement, the Smith-Waterman algorithm runs too slow to be useful for discovering homologous genes. Because of that, new methods - using heuristic approach - have been focused on reducing computation time at the expense of accuracy. Since only exact algorithms guarantee optimal alignments, there lacks a method for verifying results on genome scale.

In the last few years the popularity of using CUDA architecture based graphical processors (GPUs) has intensified work on accelerating the Smith-Waterman algorithm. The majority of reports presented GPU implementations for protein database searches (Manavski and Valle, 2008; Munekawa *et al.*, 2008; Ling *et al.*, 2009; Liu *et al.*, 2009, 2010; Hains *et al.*, 2011; Zhang *et al.*, 2011). These implementations do not reduce memory consumption and therefore limit the input length of sequences. To the best of our knowledge, the only pairwise GPU implementation of the Smith-Waterman algorithm tested on the chromosome level is CUDAlign (E. F. O. Sandes and A. C. de Melo, 2010; de O. Sandes and A. C. M. a. de Melo, 2011; E. F. D. O. Sandes and A. C. M. A. De Melo, 2013) However their application intensively uses disk space and it is not publicly available.

In this paper we present our GPU implementation of the Smith-Waterman algorithm for long sequences that does not require any additional disk space and can be run on multiple cards.

## 2 METHODS

Smith-Waterman algorithm parallelization is most commonly done by solving the elements on the matrix anti-diagonal at the same time. This method is also known as the wavefront method. Method given in (E. F. D. O. Sandes and A. C. M. A. De Melo, 2013) uses the wavefront method and adapts it for usage with CUDA architecture. The solving matrix is divided into cell blocks and each block is divided into two phases, because of synchronization purposes. Each of the cell blocks is solved by one CUDA block and is considered as a single solving unit. In this way $B*T$ cells can be calculated in the same time, where $B$ is the number of CUDA blocks and $T$ is the number of CUDA threads. This is accomplished in memory complexity of $O(n + 9 * m)$, where $n$ and $m$ are lengths of longer and shorter sequences, respectively. Output of this phase of solving is the score and the endpoint of the alignment. As the left picture in Figure 1 shows the best local alignment could include only upper-half, only lower-half, or both halves of the matrix.

*To whom correspondence should be addressed.

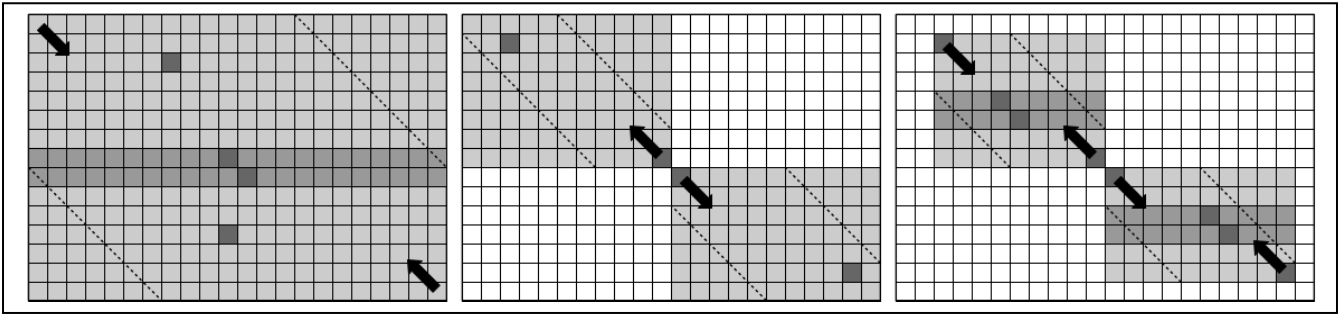

**Figure 1 SW# done with two CUDA cards.** Left, middle and right pictures show solving, finding alignment endpoints and reconstruction phases, respectively. Grey, darker grey and darkest cells represent executable area, middle cells and maximum score cells, respectively. The maximum score at the first phase can be in achieved in the upper part, in the lower part or in the middle as the sum of scores of neighboring cells. The black arrows show direction of execution, while dashed lines represent prunable area.

Wavefront phase can be optimized by not solving all of the cells of the solving matrix. As shown in (E. F. D. O. Sandes and A. C. M. A. De Melo, 2013) there can exist cells that cannot in any case achieve the optimal score. Prunable cells are cells which in any case till the end of the solving matrix cannot achieve score as high as the current optimal score. This idea is used in the wavefront method to reduce the number of blocks that need to be solved. Pruning method can reduce the number of blocks as high as 50% (E. F. D. O. Sandes and A. C. M. A. De Melo, 2013).

Next step of the solving procedure is to find the starting points of the alignment (Figure 1, picture in the middle). This is done by solving the modified Smith-Waterman algorithm on reverse subsequences which start from the found endpoint. The differences from the original Smith-Waterman algorithm is that all of the borders of the matrix are initialized to infinity to assure that the alignment starts in the endpoint and that the scores can drop below zero to assure the found startpoint is connected to the endpoint. The same wavefront method of parallelization is used as in the phase of finding the endpoint. In this phase modified banded algorithm (Ukkonen, 1985) can be used. The maximum deformation t can be calculated as the known alignment score dived by the highest substitution score. The maximum length m' represents the maximum possible cells of one sequence that alignment can contain and it is calculated as the sum of the length of longer sequence and longest possible gap, which is calculated from the score. If the lengths of the sequences are m and n, where $m > n$, $m'$ can be calculated as $min(n + (n - t) / ge, m)$, where $ge$ is the gap extension penalty. Padding $p$ (Ukkonen, 1985) can then be calculated as $ceil(0.5 * (2 * n - t - m'))$. This observation is done by calculating the maximum score a cell can reach and comparing it to the found score. Further on the banded algorithm for cell pruning is used and all the cell blocks that don't contain any of diagonals between $-p$ and $p + (m - n)$ are pruned. Memory complexity of this procedure is $O(n + 9 * m)$, where $n$ and $m$ are lengths of the sequences entering this phase, which can be considerably shorter than the original sequences. Outputs of this stage of are the ending points of the alignment.

The reconstruction phase considers only the cells between the start and the endpoint (Figure 1, the right picture). Modified Myers-Miller algorithm is applied (Myers and Miller, 1988). This phase is done in parallel by both the CPU and the GPU. GPU part of the algorithm starts the Myers-Miller algorithm on the halves of the matrix. If multiple cards are available they are used on the halves. Every part of the matrix is solved with the wavefront method of the

Needleman-Wunsch algorithm (Needleman and Wunsch, 1970). Parallelization of the Needleman-Wunsch algorithm is done in the same way as the Smith-Waterman algorithm. Exact banded algorithm can be applied. The padding is calculated on the whole matrix and given to the solving halves for pruning. The maximum edit distance $t$ is calculated as $max(m, n) - score$, where $m$ and $n$ are the lengths of the sequences between the startpoint and the endpoint and the score is the alignment score. The Myers-Miller algorithm is applied in this way recursively. Difference from the original algorithm is that is stops as soon as the solving submatrix size drops bellow defined boundaries. When the size of solving submatrix is small enough GPU part sends it to the CPU part. CPU part performs the Needleman-Wunsch alignment on the solving submatrix. As the score of the submatrix is known, banded algorithm is applied. When the GPU part is over, and CPU solves all of the submatrices, alignments of the submatrices are joined in the complete alignment. Main advantage of this method is that the memory complexity is the complexity of the wavefront method $O(n + 9 * m)$, where $n$ and $m$ are lengths of the sequences between the startpoint and the endpoint. Output of this phase is the alignment of the input sequences.

The Smith-Waterman algorithm and its CUDA implemented wavefront method can easily be divided in solving the two subproblems with at least two GPU cards. Applying the first step of Myers-Miller algorithm the endpoint and the score of the upper half, the endpoint and the score of the reverse lower part of the matrix are calculated as well as the last row of the upper and lower part. Applying ideas from Myers-Miller algorithm we get the middle score and the midpoint from the two calculated rows. In this stage we have three scores, in the lower part, in the upper part and at the midpoint. There are three situations that could happen:

a) Maximum score is in the lower part. The bottom endpoint is the startpoint of the alignment. End point is found in the same way as the startpoint in the single card method. Reconstruction is then also done as the single card method.

b) Maximum score is in the upper part: Method is equal to the single card method.

c) Maximum score is at the midpoint: In parallel two algorithms are performed. The maximum score is the sum, at the midpoint, of scores from both lower and upper parts of the matrix. At first lower and upper parts are calculated separately and then they are concatenated.

## 3 RESULTS

Application run times are displayed in Table 1. Measurements are done using two different GPU cards and CPU Intel® Core™2 Quad Processor Q6600. The GTX 570 GPU card we performed measurements is fairly similar to GTX 560 used in (E. F. D. O. Sandes and A. C. M. A. De Melo, 2013). We presented result achieved with maximum of two GPU cores. The maximum speed up could be achieved using four GPU cards in the case of aligning both strands. In this case the total running times for both strands are similar to results presented in Table 1 for one strand. Compared with presented results for CUDAlign, SW# is slower, using only one GPU card, in cases when the alignment is very long. However, in difference to CUDAlign it does not use any additional disk space, can use multiple cards and the code is publicly available.

This application can be executed on Windows, Linux and Mac OS operating systems. As this application is a part of a bigger library only the swhrapn, swsharpp and swsharpnc executables are relevant.

**Table 1. SW# and CUDAlign comparison for different NVIDIA GPU cards and sequence lengths.** CPU results are presented only for shorter sequences. CUDAlign results are taken from the original paper.

| Sequences size | CPU | GTX 560 CUDAlign | GTX 570 SW# | GTX 690* SW# |
|---|---|---|---|---|
| 162Kb×172Kb | 660s | 2.1s | 1.5s | 2.6s |
| 0.5Mb×0.5Mb | 9090s | 11.8s | 9.4s | 5.8s |
| 3.1Mb×3.3Mb | | 367s | 296s | 119s |
| 7.1Mb×5.2Mb | | 1321s | 1063s | 422s |
| 23Mb×25Mb | | 19757s | 15979s | 6369s |
| 59Mb×24Mb | | 47123s | 40359s | 16263s |
| 33Mb×47Mb | | 30369s | 59228s | 23614s |

*card with 2 GPU cores

## ACKNOWLEDGEMENTS

The authors would like to thank Gloria Fuentes, Nenad Malenica, Ivan Dokmanic and Ivana Mihalek for valuable comments on the manuscript.

*Funding*: This work was supported by Biomedical Research Council of A*STAR, Singapore and by Ministry of Education Science and Sports of Republic of Croatia [036-0362214-1987] to M.Š.